\documentclass{llncs}
\usepackage{amssymb}
\usepackage{amsmath}
\usepackage{texdraw}
\usepackage[dvipdf]{graphicx}

\begin{document}

\title{\bf Candidate One-Way Functions and One-Way Permutations Based on Quasigroup String Transformations}
\author{ D. Gligoroski\inst{1,}\inst{2}}

\institute{Centre for Quantifiable Quality of Service in
Communication Systems, Norwegian University of Science and
Technology, O.S.Bragstads
plass 2E, N-7491 Trondheim, NORWAY \\
%
\and University - ``Ss Cyril and Methodius'', Faculty of Natural
Sciences and Mathematics, Institute of Informatics,
P.O.Box 162, 1000 Skopje, Republic of MACEDONIA \\
\email{gligoroski@yahoo.com}}

\maketitle
\date\

\begin{abstract}
In this paper we propose a definition and construction of a new
family of one-way candidate functions ${\cal R}_N:Q^N \rightarrow
Q^N$, where $Q=\{0,1,\ldots,s-1\}$ is an alphabet with $s$
elements. Special instances of these functions can have the
additional property to be permutations (i.e. one-way
permutations). These one-way functions have the property that for
achieving the security level of $2^n$ computations in order to
invert them, only $n$ bits of input are needed. The construction
is based on quasigroup string transformations. Since quasigroups
in general do not have algebraic properties such as associativity,
commutativity, neutral elements, inverting these functions seems
to require exponentially many readings from the lookup table that
defines them (a Latin Square) in order to check the satisfiability
for the initial conditions, thus making them natural candidates
for one-way functions. \footnote{This work was carried out during
the tenure of an ERCIM fellowship of D. Gligoroski visiting the
Centre Q2S - Centre for Quantifiable Quality of Service in
Communication Systems at the Norwegian University of Science and
Technology - Trondheim, Norway.}

\vspace{0.5cm} {\it Key words: one-way functions, one-way
permutations, quasigroup string transformations}


\end{abstract}

\newpage
\section{Introduction}%

Almost all known and well established one-way functions and
one-way permutations in modern cryptography are based on
intractable problems from number theory or closely related
mathematical fields such as theory of finite fields, sphere
packing or coding theory. For example, the discrete logarithm
problem modulo a large randomly generated prime number is the
Diffie-Helman proposal in \cite{DiffieHelman} for one-way
permutations, quadratic residuosity is Goldwasser and Micali
proposal in \cite{GoldwasserMicali} and RSA is an one-way
permutation candidate based on the difficulty of factoring a
number that is a product of two large prime numbers proposed by
Rivest, Shamir and Adleman in \cite{RSA}. There are also some
one-way functions candidates based on sphere-packing problems and
coding theory such as the proposals from Goldreich, Krawczyk and
Luby in \cite{GoldreichKrawczykLuby}. Constructing one-way
functions based on the subset sum problem have been proposed by
Impagliazzo and Naor in \cite{ImpagliazzoNaor}. As far as we know,
the only attempt to construct a one-way function that is
completely defined by combinatorial elements is the proposal of
Goldreich in \cite{Goldreich}. The proposal is based on the
combinatorial field of Expander Graphs.

In this paper we construct a new family of one-way functions and
one-way permutations defined on a finite set
$Q=\{0,1,\ldots,s-1\}$ with $s$ elements. The construction is
based on the theory of quasigroups, and quasigroup string
transformations. Our approach in opposite to other approaches,
with an exception of \cite{Goldreich} is completely based on a
mathematical field not closely related to the field of number
theory. By some of their properties (such as speed of computation,
security level of inversion) quasigroup one-way functions
outperform all currently known one-way candidate functions.


\section{Preliminaries}

Here we give a brief overview of quasigroups and quasigroup string
transformations and more detailed explanation the reader can find
in \cite{DenKeed} and \cite{Quasi1}.

\begin{definition}
A quasigroup $(Q,*)$ is a groupoid, i.e. a set $Q$ with a binary
operation $*:Q\times Q \to Q$, satisfying the law
\begin{equation}(\forall u,v\in Q)(\exists!\ x,y\in Q) \quad u*x = v\ \& \ y*u
= v. \\ \label{QuasigroupDef}\end{equation}  \end{definition}


If $Q$ is a finite set then the main body of the multiplication
table of the quasigroup is a Latin Square over the set $Q$. A
Latin Square over $Q$ is a $|Q|\times |Q|$-matrix such that each
row and column is a permutation of $Q$ \cite{DenKeed}.

Next we define the basic quasigroup string transformation called
$e$--transformation:

\begin{definition} A quasigroup $e$--transformation of a string $A=(a_0,a_1,\dots,a_{N-1}) \in Q^N$
with a leader $l \in Q$ is the function $e_l:Q\times Q^N
\rightarrow Q^N$ defined as $B=e_l(A)$ where
$A=(a_0,a_1,\dots,a_{N-1})$, $B=(b_0,b_1,\dots,b_{N-1})$, $l\in Q$
and
     \begin{equation}b_i: = \left\{\begin{array}{ll}
        l*a_0,  & i = 0 \\
        b_{i-1}*a_i, & 1 \le i \le N-1
     \end{array}\right.\label{ETransformation}\end{equation}
\end{definition}

\begin{figure}[h]
\begin{center}

\unitlength=1cm
\begin{picture}(10,1.5)(-5,-0.5)
\thinlines \put(-4,0){\line(1,0){8}} \thicklines
\put(-3,-1){\line(0,1){2}} \thinlines
\multiput(-2,-1)(1,0){2}{\line(0,1){2}}
\multiput(2,-1)(1,0){3}{\line(0,1){2}}

\put(-2.5,0.5){$a_0$} \put(-1.5,0.5){$a_1$} \put(0.5,0.5){$\dots$}
\put(2.15,0.5){$a_{N-2}$} \put(3.20,0.5){$a_{N-1}$}

\put(-3.5,-0.65){$l$} \put(-2.5,-0.65){$b_0$}
\put(-1.5,-0.65){$b_1$} \put(0.5,-0.65){$\dots$}
\put(2.15,-0.65){$b_{N-2}$} \put(3.20,-0.65){$b_{N-1}$}

\multiput(-3.35,-0.5)(1,0){3}{\vector(1,1){0.85}}
\multiput(1.65,-0.5)(1,0){2}{\vector(1,1){0.85}}
\multiput(-2.35,0.35)(1,0){2}{\vector(0,-1){0.65}}
\multiput(2.65,0.35)(1,0){2}{\vector(0,-1){0.65}}

\end{picture}
\end{center}
\caption{Graphical representation of the $e$--transformation of a
string $A=(a_0,a_1,\dots,a_{N-1})$.} \label{FigETransformation}
\end{figure}
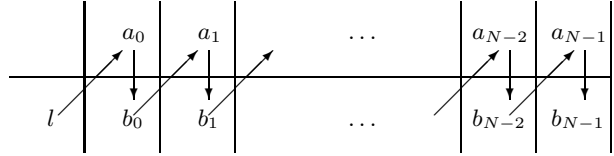

For better understanding the graphical representation of the
$e$--transformation is shown on Fig. \ref{FigETransformation}.

\begin{example}
Let  $Q=\{0,1,2,3\}$ and let the quasigroup $(Q,*)$ be given by
the multiplication scheme in Table \ref{ExampleQuasigroups4}.

\begin{table}
\begin{center}
\begin{tabular}{c|rrrrr}
\multicolumn{5}{c}{}\\
    $*$ &  0 & 1 & 2 & 3\\
\hline
       0    & 2 & 1 & 0 & 3\\
       1    & 3 & 0 & 1 & 2\\
       2    & 1 & 2 & 3 & 0\\
       3    & 0 & 3 & 2 & 1\\
\end{tabular}
\end{center}
\caption{Quasigroup $(Q,*)$}
 \label{ExampleQuasigroups4}
\end{table}

Consider the string  $A = 1\ 0 \ 2 \ 1 \ 0 \ 0 \ 0 \ 0 \ 0 \ 0 \ 0
\ 0 \ 0 \ 1 \ 1 \ 2 \ 1 \ 0 \ 2 \ 2 \ 0 \ 1 \ 0 \ 1 \ 0 \ 3 \ 0 \
0 $ and let us choose the leader $l=0$. Then by the
$e$--transformation $e_{0}(A)$ we will obtain the following
transformed string:

$e_{0}(A) = 1\ 3\ 2\ 2\ 1\ 3\ 0\ 2\ 1\ 3\ 0\ 2\ 1\ 0\ 1\ 1\ 2\ 1\
1\ 1\ 3\ 3\ 0\ 1\ 3\ 1\ 3\ 0 .$

The four consecutive applications of the $e$-transformation
$e_{0}$ on $A$ are represented in Table \ref{example e e'}.
\begin{table}[h]
\begin{center}
\begin{tabular}{c|l}
    &  $1\ 0 \ 2 \ 1 \ 0 \ 0 \ 0 \ 0 \ 0 \ 0
\ 0 \ 0 \ 0 \ 1 \ 1 \ 2 \ 1 \ 0 \ 2 \ 2 \ 0 \ 1 \ 0 \ 1 \ 0 \ 3 \
0 \ 0 = A$ \\ \hline
       0    & $1\ 3\ 2\ 2\ 1\ 3\ 0\ 2\ 1\ 3\ 0\ 2\ 1\
0\ 1\ 1\ 2\ 1\ 1\ 1\ 3\ 3\ 0\ 1\ 3\ 1\ 3\ 0 = e_{0}(A)$\\
       0    &$1\ 2\ 3\ 2\ 2\ 0\ 2\ 3\ 3\ 1\ 3\ 2\ 2\ 1\ 0\ 1\ 1\ 2\ 2\ 2\ 0\ 3\ 0\ 1\ 2\ 2\ 0\ 2 = e_{0}( e_{0}(A) )$ \\
       0    & $1\ 1\ 2\ 3\ 2\ 1\ 1\ 2\ 0\ 1\ 2\ 3\ 2\ 2\ 1\ 0\ 1\ 1\ 1\ 1\ 3\ 1\ 3\ 3\ 2\ 3\ 0\ 0= e_{0}( e_{0}( e_{0}(A) ) )$\\
       0    & $1\ 0\ 0\ 3\ 2\ 2\ 2\ 3\ 0\ 1\ 1\ 2\ 3\ 2\ 2\ 1\ 0\ 1\ 0\ 1\ 2\ 2\ 0\ 3\ 2\ 0\ 2\ 1= e_{0}( e_{0}( e_{0}( e_{0}(A) ) ) )$\\
\end{tabular}\\
\end{center}
\caption{Four consecutive $e$-transformations of $A$ with leader
$0$.}
 \label{example e e'}
\end{table}
\end{example}



If we have a string of leaders, we can apply consecutive
$e$--transformations on a given string, as a composition of
$e$--transformations. That is defined by the following definition:

\begin{definition} A quasigroup $E$--transformation of a string $A=(a_0,a_1,\dots,a_{N-1}) \in Q^N$
with a string of $K$ leaders $\mathbf{L}=(l_0,l_1,\dots,l_{K-1})
\in Q^K$ is the function $E_{\mathbf{L},K}:Q^K \times Q^N
\rightarrow Q^N$ defined as $B=E_{\mathbf{L},K}(A)$ where
$A=(a_0,a_1,\dots,a_{N-1})$, $B=(b_0,b_1,\dots,b_{N-1})$ and
     \begin{equation}
     B=e_{l_{K-1}}(\ e_{l_{K-2}}(\ \ldots e_{l_1}(\ e_{l_0}(A)\ )
     \dots \ ) \ )
     \label{EETransformation}
     \end{equation}
\end{definition}

\begin{definition} Quasigroup single reverse string transformation is the function ${\cal R}_1:Q^N
\rightarrow Q^N$ defined as $$B={\cal
R}_1(A)=E_{\overline{A},N}(A)=e_{a_{N-1}}(\dots(e_{a_1}(e_{a_0}(A))))$$
where $A=(a_0,a_1,\dots,a_{N-1})$ and $B=(b_0,b_1,\dots,b_{N-1})$.
\end{definition}

\begin{definition} Quasigroup double reverse transformation is the function ${\cal R}_2:Q^N
\rightarrow Q^N$ defined as
$$B={\cal
R}_2(A)=E_{\overline{A}\overline{A},2N}(A)=e_{a_{N-1}}(\dots(e_{a_1}(e_{a_0}(e_{a_{N-1}}(\dots(e_{a_1}(e_{a_0}(A))))$$
where $A=(a_0,a_1,\dots,a_{N-1})$ and $B=(b_0,b_1,\dots,b_{N-1})$.
\end{definition}

\begin{example}
Let quasigroup $(Q,*)$ be given by the multiplication scheme in
Table \ref{ExampleQuasigroups4}. Consider the string  $A = 0\ 1 \
2 \ 3 \ 0 $. Then by the transformation ${\cal
R}_1(A)=E_{\overline{A},5}(A)$ we will obtain the following
transformed string: ${\cal R}_1(A) = 0\ 0\ 1\ 0\ 3$ and by the
transformation ${\cal R}_2(A)=E_{\overline{A}\overline{A},10}(A)$
we will obtain the following transformed string: ${\cal R}_2(A) =
0\ 3\ 2\ 0\ 2$. The calculation's steps are given in Table
\ref{exampleR1R2}.

\begin{table}
\begin{center}
\begin{tabular}{ccc}
\begin{tabular}{c|rrrrrl}
\multicolumn{6}{c}{}\\
            & 0 & 1 & 2 & 3 & 0 & $=A$\\
\hline
       0    & 2 & 2 & 3 & 1 & 3 &\\
       3    & 2 & 3 & 1 & 0 & 3 &\\
       2    & 3 & 1 & 0 & 2 & 0 &\\
       1    & 2 & 2 & 1 & 1 & 3 &\\
       0    & {\bf 0} & {\bf 0} & {\bf 1} & {\bf 0} & {\bf 3} & $={\cal R}_1(A)$\\
\end{tabular}
& \qquad \qquad \qquad &
\begin{tabular}{c|rrrrrl}
\multicolumn{6}{c}{}\\
            & 0 & 1 & 2 & 3 & 0 & $=A$\\
\hline
       0    & 2 & 2 & 3 & 1 & 3 &\\
       3    & 2 & 3 & 1 & 0 & 3 &\\
       2    & 3 & 1 & 0 & 2 & 0 &\\
       1    & 2 & 2 & 1 & 1 & 3 &\\
       0    & 0 & 0 & 1 & 0 & 3 &\\
       0    & 2 & 1 & 0 & 2 & 0 &\\
       3    & 2 & 2 & 1 & 1 & 3 &\\
       2    & 3 & 2 & 2 & 2 & 0 &\\
       1    & 2 & 3 & 2 & 3 & 0 &\\
       0    & {\bf 0} & {\bf 3} & {\bf 2} & {\bf 0} & {\bf 2} & $={\cal R}_2(A)$\\
\end{tabular}
\end{tabular}
\end{center}
\caption{${\cal R}_1(A)$ and ${\cal R}_2(A)$ transformation of the
string $A = 0\ 1 \ 2 \ 3 \ 0 $.}
 \label{exampleR1R2}
\end{table}

\end{example}

\section{One-wayness from the lookup table point of view}

Both ${\cal R}_1$ and ${\cal R}_2$ are serious candidates for
one-way functions, with the difference that the number of
computations to invert ${\cal R}_1$ is
$O(s^{\left\lfloor\frac{N}{3}\right\rfloor})$ and to invert ${\cal
R}_2$ it is $O(s^{N})$. In the following two theorems we will
prove these claims from a perspective of the lookup table (Latin
Square) that defines the used quasigroup $(Q,*)$. We will discuss
later in this section the reasons for this approach.

\begin{theorem}
If the quasigroup $(Q,*)$ is non-associative and non-commutative,
then the number of computations based only on the lookup table
that defines the quasigroup $(Q,*)$ in order to find the preimage
for the function ${\cal R}_1:Q^N \rightarrow Q^N$ is
$O(s^{\left\lfloor\frac{N}{3}\right\rfloor})$.
\end{theorem}
{\bf Proof:} Let $B=(b_0,b_1,\dots,b_{N-1})$ be given. The goal is
to find a string $A=(a_0,a_1,\dots,a_{N-1})$ that satisfies the
equality $B=E_{\overline{A},N}(A)=E_{(a_{N-1},\, a_{N-2},\dots ,\,
a_1, \, a_0),N}(A)$. Further, because the final values of the
string $B$ are obtained after $N$ consecutive operations $e_{a_j}$
we will use the following notation:
$B^{(i)}=e_{a_{N-i}}(B^{(i-1)})=(b_0^{(i)},b_1^{(i)},\dots,b_{N-2}^{(i)},b_{N-1}^{(i)})$
for $i=\{1,\dots,N-1\}$, and $B^{(0)}=A$, $B^{(N)}\equiv B$.

\begin{table}[h]
\label{table1a}
\begin{center}
\begin{tabular}{c|ccccc}
 & $?$ & $?$ &\dots & \dots &$?$\\
\hline
\\
$?$ & $?$ & $?$ &\dots & \dots &$b_{N-1}^{(1)}$ \\
\\
$?$ & $?$ & $?$ &\dots & \dots &$b_{N-1}^{(2)}$ \\
\\
$\vdots$ & $\vdots$ & $\vdots$ & $\ddots$ & . &$\vdots$\\
\\
$?$ & $?$ & $?$ & $b_2^{(N-2)}$ &  \dots & $b_{N-1}^{(N-2)}$\\
\\
$?$ & $?$ & $b_1^{(N-1)}$ &\dots &  \dots & $b_{N-1}^{(N-1)}$
\\
\hline $?$ & $b_0^{(N)}$ & $b_1^{(N)}$ & \dots &  \dots
&$b_{N-1}^{(N)}$
\end{tabular}
\end{center}
\caption{Initial table obtained from the values of
$B=(b_0,b_1,\dots,b_{N-1})$ before making any guess for the values
of $A=(a_0,a_1,\dots,a_{N-1})$.} \label{Table4}
\end{table}

\begin{table}[h]
\begin{center}
\begin{tabular}{ccc}
\begin{tabular}{c|ccccc}
 & $a_0$ & $?$ &\dots & $?$ &$a_{N-1}$\\
\hline
\\
$a_{N-1}$ & $b_0^{(1)}$ & $?$ &\dots & $b_{N-2}^{(1)}$ & $b_{N-1}^{(1)}$ \\
\\
$?$ & $?$ & $?$ &\dots & $b_{N-2}^{(2)}$ & $b_{N-1}^{(2)}$ \\
\\
$\vdots$ & $\vdots$ & $\vdots$ & . & . &$\vdots$\\
\\
$?$ & $?$ & $b_1^{(N-2)}$ & $b_2^{(N-2)}$ &  \dots & $b_{N-1}^{(N-2)}$\\
\\
$?$ & $b_0^{(N-1)}$ & $b_1^{(N-1)}$ &\dots &  \dots &
$b_{N-1}^{(N-1)}$
\\
\hline $a_0$ & $b_0^{(N)}$ & $b_1^{(N)}$ & \dots &  \dots
&$b_{N-1}^{(N)}$
\end{tabular}

  &

   \qquad  \qquad

  &

\begin{tabular}{c|ccccc}
 & $a_0$ & $a_1$ &\dots & $a_{N-2}$ & $a_{N-1}$\\
\hline
\\
$a_{N-1}$ & $b_0^{(1)}$ & $b_1^{(1)}$ &\dots & $b_{N-2}^{(1)}$ & $b_{N-1}^{(1)}$ \\
\\
$a_{N-2}$ & $b_0^{(2)}$ & $b_1^{(2)}$ &\dots & $b_{N-2}^{(2)}$ & $b_{N-1}^{(2)}$ \\
\\
$\vdots$ & $\vdots$ & $\vdots$ & . & . &$\vdots$\\
\\
$?$ & $?$ & $b_1^{(N-3)}$ & $b_2^{(N-3)}$ &  \dots & $b_{N-1}^{(N-3)}$\\
\\
$?$ & $b_0^{(N-2)}$ & $b_1^{(N-2)}$ & $b_2^{(N-2)}$ &  \dots & $b_{N-1}^{(N-2)}$\\
\\
$a_1$ & $b_0^{(N-1)}$ & $b_1^{(N-1)}$ &\dots &  \dots &
$b_{N-1}^{(N-1)}$
\\
\hline $a_0$ & $b_0^{(N)}$ & $b_1^{(N)}$ & \dots &  \dots
&$b_{N-1}^{(N)}$
\end{tabular}  \\

 & & \\

$\begin{array}{c}\mbox{a. Completing the table when}\\
\mbox{the value of $a_0$ is guessed.}\end{array} $ & &
$\begin{array}{c}\mbox{b. Completing the table when}\\
\mbox{the values of $a_0$ and $a_1$ are guessed.}\end{array} $
\end{tabular}
\end{center}
\caption{} \label{Table5}
\end{table}

Since the quasigroup $(Q,*)$ is non-associative and
non-commutative, the composition of $e$--transformations is fixed
and it can not be changed (this is not the case if the quasigroup
is commutative or associative). Thus, to solve the inverse task in
fact we have to fill in the scheme in the Table \ref{Table4}, from
bottom up using the properties of the quasigroup operation $*$. As
a matter of fact due to the properties of quasigroup operation $*$
this scheme can be partially completed without guessing any value
of $A$.
Namely, from the equation $b_i^{(N)}*x=b_{i+1}^{(N)}$ we can
calculate $x= b_{i+1}^{(N-1)}$ for $0\le i\le N-1$, then from
$b_i^{(N-1)}*y=b_{i+1}^{(N-1)}$ we can calculate $y=
b_{i+1}^{(N-2)}$ for $1\le i\le N-1$, and so on up to the first
row of the table, where we can calculate the value of
$b_{N-1}^{(1)}$.

Now, by knowing (or by guessing) the value of $a_0$ that range
among $s$ possible values we can find value $b_0^{(N-1)}$, from
which we can find the other values in the scheme of Table
\ref{Table5}a, together with the value of $a_{N-1}$.

If we continue with choosing $a_1$ from all possible $s$ values we
will obtain a new value for $a_{N-2}$. Next, with every choice of
$a_i,\, 2\le i \le \frac{N}{2}$ we will obtain also the values for
$a_{N-i-1}$, and by knowing that we will be in a position to
complete the upper left corner of the scheme (see Table
\ref{Table5}b). The intersection of the lower completed and the
upper completed part is for
$\left\lfloor\frac{N}{3}\right\rfloor$. So by choosing
$\left\lfloor\frac{N}{3}\right\rfloor$ values we will obtain other
values of the string $A$. Now, we can check whether we have made
the right choice for $a_0,\,
a_1,\dots,a_{\left\lfloor\frac{N}{3}\right\rfloor}$ or not.
Therefore, the complexity of inversion of ${\cal R}_1$ only by
using the lookup definition of the quasigroup $(Q,*)$ is
$O(s^{\left\lfloor\frac{N}{3}\right\rfloor})$. $\square$

\begin{theorem}
If the quasigroup $(Q,*)$ is non-associative and non-commutative,
then the number of computations based only on the lookup table
that defines the quasigroup $(Q,*)$ in order to find the preimage
for the function ${\cal R}_2:Q^N \rightarrow Q^N$ is $O(s^N)$.
\end{theorem}
{\bf Proof:} The proof is similar to the proof for the function
${\cal R}_1$ except that now there is no intersection in the
process of completing the scheme until the last guess for
$a_{N-1}$ is made. Therefore we have to make a guess for all $N$
values $a_0, a_1, \ldots, a_{N-1}$ and thus the complexity of
inverting the function ${\cal R}_2$ only by using the lookup
definition of the quasigroup $(Q,*)$ is $O(s^N)$. $\square$

From previous two theorems we can make the following conjecture:

\begin{conjecture}
${\cal R}_1$ and ${\cal R}_2$ are one-way functions.
\end{conjecture}

To support Conjecture 1 we would like to stress that the used
quasigroup $(Q,*)$ in general will not have any algebraic property
such as commutativity, associativity, neutral elements etc. Thus,
the only possible way to deal with the problem of inversion of
these functions is to look at the lookup table (or Latin Square)
that defines the quasigroup $(Q,*)$.

\vspace{-0.5cm}
\begin{table}[h!]
\begin{center}
\begin{tabular}{c|ccccc}
$\mathcal{R}_N(A)$  & $a_0$ & $a_1$ &$\dots$ & $a_{N-2}$ & $a_{N-1}$\\
\hline $\mathbf{L} \left\{ \begin{array}{c}
l_0 \\
l_1 \\
\vdots\\
l_{P(N)}
\end{array}\right. $        &   $\begin{array}{c}
. \\
. \\
\vdots\\
.
\end{array}$
   &    $\begin{array}{c}
. \\
. \\
\vdots\\
.
\end{array}$
  &   $\dots$  &     $\begin{array}{c}
. \\
. \\
\vdots\\
.
\end{array}$
     &     $\begin{array}{c}
. \\
. \\
\vdots\\
.
\end{array}$
    \\

$\overline{A} \left\{ \begin{array}{c}
a_{N-1} \\
a_{N-2} \\
\vdots\\
a_{0}
\end{array}\right. $  &   $\begin{array}{c}
. \\
. \\
\vdots\\
.
\end{array}$   &    $\begin{array}{c}
. \\
. \\
\vdots\\
.
\end{array}$  &  \dots & $\begin{array}{c}
. \\
. \\
\vdots\\
.
\end{array}$ & $\begin{array}{c}
. \\
. \\
\vdots\\
.
\end{array}$ \\
$\overline{A} \left\{ \begin{array}{c}
a_{N-1} \\
a_{N-2} \\
\vdots\\
a_{0}
\end{array}\right. $  &   $\begin{array}{c}
. \\
. \\
\vdots\\
b_0
\end{array}$   &    $\begin{array}{c}
. \\
. \\
\vdots\\
b_1
\end{array}$  &  $\begin{array}{c}
\dots \\
\dots \\
\vdots\\
\dots
\end{array}$ & $\begin{array}{c}
. \\
. \\
\vdots\\
b_{N-2}
\end{array}$ & $\begin{array}{c}
. \\
. \\
\vdots\\
b_{N-1}
\end{array}$ \\
\end{tabular}
\end{center}
\caption{Schematic representation of the process of computation of
the function $\mathcal{R}_N$.} \label{RN}
\end{table}

\vspace{-0.5cm} Next we will use the function ${\cal R}_2$ as a
core for defining a family of one-way function candidates. The
idea is that before applying the function ${\cal R}_2$ on some
string $A$ of length $N$, we would like to apply a certain number
(polynomial on $N$) of $e$--transformations with leaders that are
some constants from $Q$ or they are fixed indexes that address
certain letters of the string $A$. For that purpose we will need
the following definition:

\begin{definition}
Preprocessing string of leaders
$\mathbf{L}=\mathbf{L}_{Q,I_N,P(N)}=(l_0,l_1,\ldots,l_{P(N)})$ is
a string of length that is polynomial of $N$ and where $l_i \in Q
\cup I_N$, $Q=\{0,1,\ldots,s-1\}$ and $I_N =
\{i_0,i_1,\ldots,i_{N-1}\}$ in an index set. By convention,
$\mathbf{L}$ can be also an empty string.
\end{definition}

\begin{definition} The family $\mathcal{Q}_N $ of quasigroup one-way
functions of strings of length $N$ consists of functions ${\cal
R}_N:Q^N \rightarrow Q^N$ such that $$B={\cal
R}_N(A)=E_{\mathbf{L}\overline{A}\overline{A},P(N)+2N}(A)$$ where
$\mathbf{L}$ is defined in Deffinition 6, and $A, B \in Q^N$. By
convention, when applying the $e$--transformations with index
leader i.e. $l_j \in I_N$, then $e$--transformation have to be
applied with the leader $a_{l_j}$.
\end{definition}

For better understanding, a schematic representation of the
process of computation of the function $\mathcal{R}_N$ is given in
Table \ref{RN}.

\begin{conjecture}
The family $\mathcal{Q}_N $ is a family of one-way functions.
\end{conjecture}

\begin{example}
Let chose $N=2$ and $(Q,*)$ be as in Table
\ref{ExampleQuasigroups4}. If we interpret the elements of
$Q=\{0,1,2,3\}$ as two-bit letters $\{00,01,10,11\}$ then by
having $N=2$ we will define function
$E_{\mathbf{L}\overline{A}\overline{A},P(N)+2N}(A)$ from
$\{0,1,\ldots,15\}$ into itself. If we chose
$\mathbf{L}=(3,3,i_1,i_0)$, then
$E_{(3,3,i_1,i_0)\overline{A}\overline{A},8}(A)$ is represented in
Figure \ref{ExampleGeneral}a. Notice that the function is
permutation. On the other hand if we choose
$\mathbf{L}=(3,3,i_0,i_1)$ then we will get a function  that is
not a permutation. That is represented in Figure
\ref{ExampleGeneral}b. Particular computations for the string $01
\equiv 1$ in both cases is shown in Table \ref{example01}.

\begin{table}
\begin{center}
\begin{tabular}{ccc}
\begin{tabular}{c|ccl}
            & 0 & 1 & $\equiv 0001 \equiv 1$\\
\hline
       3    & 0 & 1 &\\
       3    & 0 & 1 &\\
       1    & 3 & 3 &\\
       0    & 3 & 1 &\\
       1    & 2 & 2 &\\
       0    & 0 & 0 &\\
       1    & 3 & 0 &\\
       0    & {\bf 3} & {\bf 0} & $\equiv 1100 \equiv 12$\\
\end{tabular}
& \qquad \qquad \qquad &
\begin{tabular}{c|ccl}
            & 0 & 1 & $\equiv 0001 \equiv 1$\\
\hline
       3    & 0 & 1 &\\
       3    & 0 & 1 &\\
       0    & 2 & 2 &\\
       1    & 1 & 1 &\\
       1    & 0 & 1 &\\
       0    & 2 & 2 &\\
       1    & 1 & 1 &\\
       0    & {\bf 1} & {\bf 0} & $\equiv 0100 \equiv 4$\\
\end{tabular}
\end{tabular}
\end{center}
\caption{Transformation of the string $A = 0\ 1 $ when
$\mathbf{L}=(3,3,i_1,i_0)$ (on the left) and
$\mathbf{L}=(3,3,i_0,i_1)$ (on the right).}
 \label{example01}
\end{table}

\end{example}

\begin{figure}

\begin{texdraw}
\def\bdot {\fcir f:0 r:0.02 }
\def\Ltext #1{\bsegment
\textref h:R v:C \htext (-0.04 0){#1} \esegment}
\def\Rtext #1{\bsegment
\textref h:R v:C \htext (+0.18 0){#1} \esegment}

\linewd 0.01 \arrowheadtype t:F

\move (0 0) \Ltext 0 \bdot \avec (1.5 -0.3) \bdot \Rtext 1 %
\move (0 -0.3) \Ltext 1 \bdot \avec (1.5 -3.6) \bdot \Rtext {12}
\move (0 -0.6) \Ltext 2 \bdot \avec (1.5 -2.1) \bdot \Rtext {7}
\move (0 -0.9) \Ltext 3 \bdot \avec (1.5 -3.0) \bdot \Rtext {10}
\move (0 -1.2) \Ltext 4 \bdot \avec (1.5 -0.9) \bdot \Rtext {3}
\move (0 -1.5) \Ltext 5 \bdot \avec (1.5 -4.2) \bdot \Rtext {14}
\move (0 -1.8) \Ltext 6 \bdot \avec (1.5 -1.5) \bdot \Rtext {5}
\move (0 -2.1) \Ltext 7 \bdot \avec (1.5 -2.4) \bdot \Rtext {8}
\move (0 -2.4) \Ltext 8 \bdot \avec (1.5 -1.8) \bdot \Rtext {6}
\move (0 -2.7) \Ltext 9 \bdot \avec (1.5 -3.3) \bdot \Rtext {11}
\move (0 -3.0) \Ltext {10} \bdot \avec (1.5 0) \bdot \Rtext {0}
\move (0 -3.3) \Ltext {11} \bdot \avec (1.5 -3.9) \bdot \Rtext {13}%
\move (0 -3.6) \Ltext {12} \bdot \avec (1.5 -1.2) \bdot \Rtext{4}
\move (0 -3.9) \Ltext {13} \bdot \avec (1.5 -2.7) \bdot \Rtext{9}
\move (0 -4.2) \Ltext {14} \bdot \avec (1.5 -0.6) \bdot \Rtext {2}
\move (0 -4.5) \Ltext {15} \bdot \avec (1.5 -4.5) \bdot \Rtext {15}%
\htext (0.3 -4.8 ){a. $(3,3,i_1,i_0)$}

\move (3.0 0) \Ltext 0 \bdot \avec (4.5 -0.3) \bdot \Rtext 1 %
\move (3.0 -0.3) \Ltext 1 \bdot \avec (4.5 -1.2) \bdot \Rtext {4}
\move (3.0 -0.6) \Ltext 2 \bdot \avec (4.5 -4.2) \bdot \Rtext {14}
\move (3.0 -0.9) \Ltext 3 \bdot \avec (4.5 -3.3) \bdot \Rtext {11}
\move (3.0 -1.2) \Ltext 4 \bdot \avec (4.5 -3.3) \bdot \Rtext {11}
\move (3.0 -1.5) \Ltext 5 \bdot \avec (4.5 -4.2) \bdot \Rtext {14}
\move (3.0 -1.8) \Ltext 6 \bdot \avec (4.5 -1.2) \bdot \Rtext {4}
\move (3.0 -2.1) \Ltext 7 \bdot \avec (4.5 -0.3) \bdot \Rtext {1}
\move (3.0 -2.4) \Ltext 8 \bdot \avec (4.5 -4.5) \bdot \Rtext {15}
\move (3.0 -2.7) \Ltext 9 \bdot \avec (4.5 -3.0) \bdot \Rtext {10}
\move (3.0 -3.0) \Ltext {10} \bdot \avec (4.5 0) \bdot \Rtext {0}
\move (3.0 -3.3) \Ltext {11} \bdot \avec (4.5 -1.5) \bdot \Rtext {5}%
\move (3.0 -3.6) \Ltext {12} \bdot \avec (4.5 -1.5) \bdot \Rtext{5}%
\move (3.0 -3.9) \Ltext {13} \bdot \avec (4.5 0) \bdot \Rtext{0}%
\move (3.0 -4.2) \Ltext {14} \bdot \avec (4.5 -3.0) \bdot \Rtext {10} %
\move (3.0 -4.5) \Ltext {15} \bdot \avec (4.5 -4.5) \bdot \Rtext {15}%
\move (4.5 -0.6) \bdot \Rtext {2}%
\move (4.5 -0.9) \bdot \Rtext {3}%
\move (4.5 -1.8) \bdot \Rtext {6}%
\move (4.5 -2.1) \bdot \Rtext {7}%
\move (4.5 -2.4) \bdot \Rtext {8}%
\move (4.5 -2.7) \bdot \Rtext {9}%
\move (4.5 -3.6) \bdot \Rtext {12}%
\move (4.5 -3.9) \bdot \Rtext {13}%
\htext (3.3 -4.8 ){a. $(3,3,i_0,i_1)$}

\end{texdraw}

\caption{Functions obtained by $\mathbf{L}$ being a.
$(3,3,i_1,i_0)$ and b. $(3,3,i_0,i_1)$}
 \label{ExampleGeneral}
\end{figure}

\section{One-way functions v.s. one-way permutations -- non-fractal v.s. fractal quasigroups}
Having defined families of one-way candidate functions, we are
interested in which case functions
$E_{\mathbf{L}\overline{A}\overline{A},P(N)+2N}(A)$ are
permutations, and when they are not. In this section we will
describe our experimental findings that give some directions for
possible mathematical answers to these questions. We hope that
this paper and the findings presented here will be sufficiently
provocative for some readers to investigate them further and
possibly give some solid mathematical explanations.

There are a lot of classifications of quasigroups of a specific
order. Two main classifications are obtained by using the
algebraic properties of the quasigroups: (1) classes of isotopic
quasigroups, which are known only for quasigroups of orders up to
10 \cite{McKay} and (2) classes of isomorphic quasigroups
\cite{DenKeed}. The importance of quasigroup classification is
noted in many papers that deal with these algebraic structures
(for example see \cite{sorge}, \cite{MarkovskiClassification}).

From the point of view of this paper, classification of
quasigroups can be done according to the nature of the one-way
functions obtained by each quasigroup.

Since the number of quasigroups increases exponentially by the
order of the quasigroup, we have made our experiments mainly for
order 4 and some of our conjectures we have tested also on
quasigroups of order 5. The total number of quasigroups of order 4
is 576. Our experiments have shown that the set of all 576
quasigroups of order 4 can be divided into two classes. One class
$\mathcal{F}$ contains 192 quasigroups and the other class
$\mathcal{NF}$ contains 384 quasigroups. If we order all
quasigroups lexicographically from 1 to 576, then the class
$\mathcal{F}$ contains the following quasigroups: $\mathcal{F}=$\{
1, 2,   3, 4, 5,   7,   9,  11, 14, 18, 21, 24,
       25,  26,  27,  28,  37,  40,  43,  46,  49,  51,  54,  57,
       60,  63,  70,  71,  77,  80,  82,  83,  92,  93, 100, 101,
      110, 111, 113, 116, 121, 126, 127, 132, 133, 138, 139, 144,
      145, 146, 147, 148, 157, 160, 163, 166, 169, 170, 171, 172,
      174, 176, 178, 179, 182, 185, 189, 192, 196, 197, 203, 206,
      212, 213, 218, 222, 223, 228, 229, 232, 234, 235, 242, 243,
      246, 252, 253, 259, 262, 263, 269, 272, 274, 275, 284, 285,
      292, 293, 302, 303, 305, 308, 314, 315, 318, 324, 325, 331,
      334, 335, 342, 343, 345, 348, 349, 354, 355, 359, 364, 365,
      371, 374, 380, 381, 385, 388, 392, 395, 398, 399, 401, 403,
      405, 406, 407, 408, 411, 414, 417, 420, 429, 430, 431, 432,
      433, 438, 439, 444, 445, 450, 451, 456, 461, 464, 466, 467,
      476, 477, 484, 485, 494, 495, 497, 500, 506, 507, 514, 517,
      520, 523, 526, 528, 531, 534, 537, 540, 549, 550, 551, 552,
      553, 556, 559, 563, 566, 568, 570, 572, 573, 574, 575,
      576\}. (By the way, the quasigroup defined in Table
      \ref{ExampleQuasigroups4} by which we have performed
      examples in this paper has the lexicographic number 355.)

From numerous experiments that we have performed, we can post the
following conjectures:
\begin{conjecture}
For any quasigroup $(Q,*) \in \mathcal{F}$ and for every natural
number $N$ there exists at least one string $\mathbf{L}$ such that
the function $E_{\mathbf{L}\overline{A}\overline{A},P(N)+2N}(A)$
is a permutation in the set $\{0,1,\ldots,2^{2 N}-1\}$.
\end{conjecture}

\begin{conjecture}
For any quasigroup $(Q,*) \in \mathcal{NF}$ and for every natural
number $N$ there is no string $\mathbf{L}$ such that the function
$E_{\mathbf{L}\overline{A}\overline{A},P(N)+2N}(A)$ is a
permutation in the set $\{0,1,\ldots,2^{2 N}-1\}$. \label{NFConj}
\end{conjecture}

The classes $\mathcal{F}$ and $\mathcal{NF}$ have another
interesting ``graphical'' property. Namely, if we take the
periodic string $01230123\ldots $, and treat every letter as a
pixel with the corresponding color, then by consecutive
application of $e$--transformations with any constant leader $l$
the set of 576 quasigroups can be divided into two classes: A
class of quasigroups that give self-similar i.e. fractal images,
and the class of quasigroups that give non self-similar images. As
an example on Figure \ref{Q4647}a we show the image obtained by
the quasigroup number 46, and on Figure \ref{Q4647}b the image
obtained by the quasigroup number 47.

\begin{table}
\begin{center}
\begin{tabular}{cc}
\includegraphics[width=2.5in]{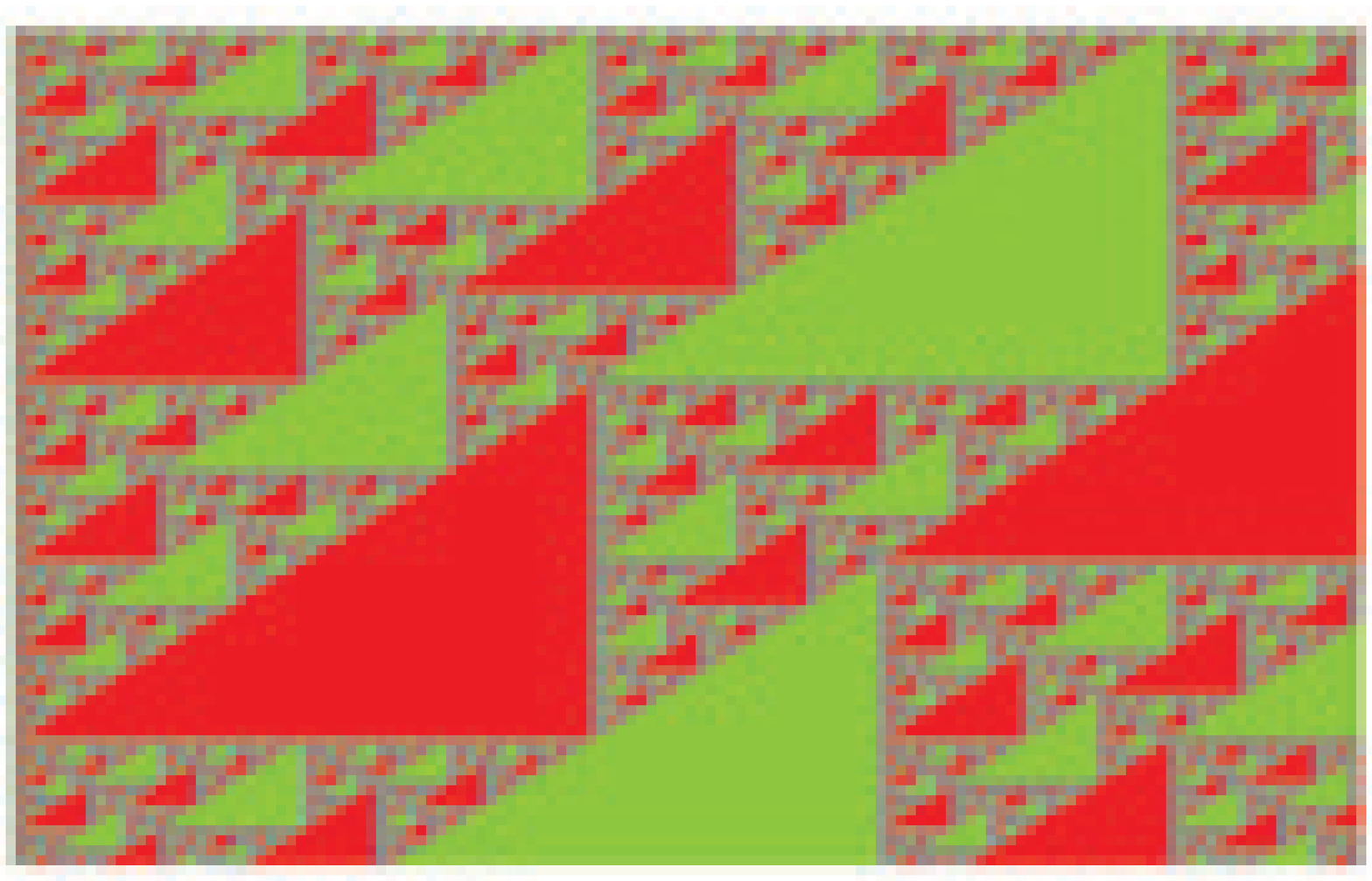}

&

\includegraphics[width=2.5in]{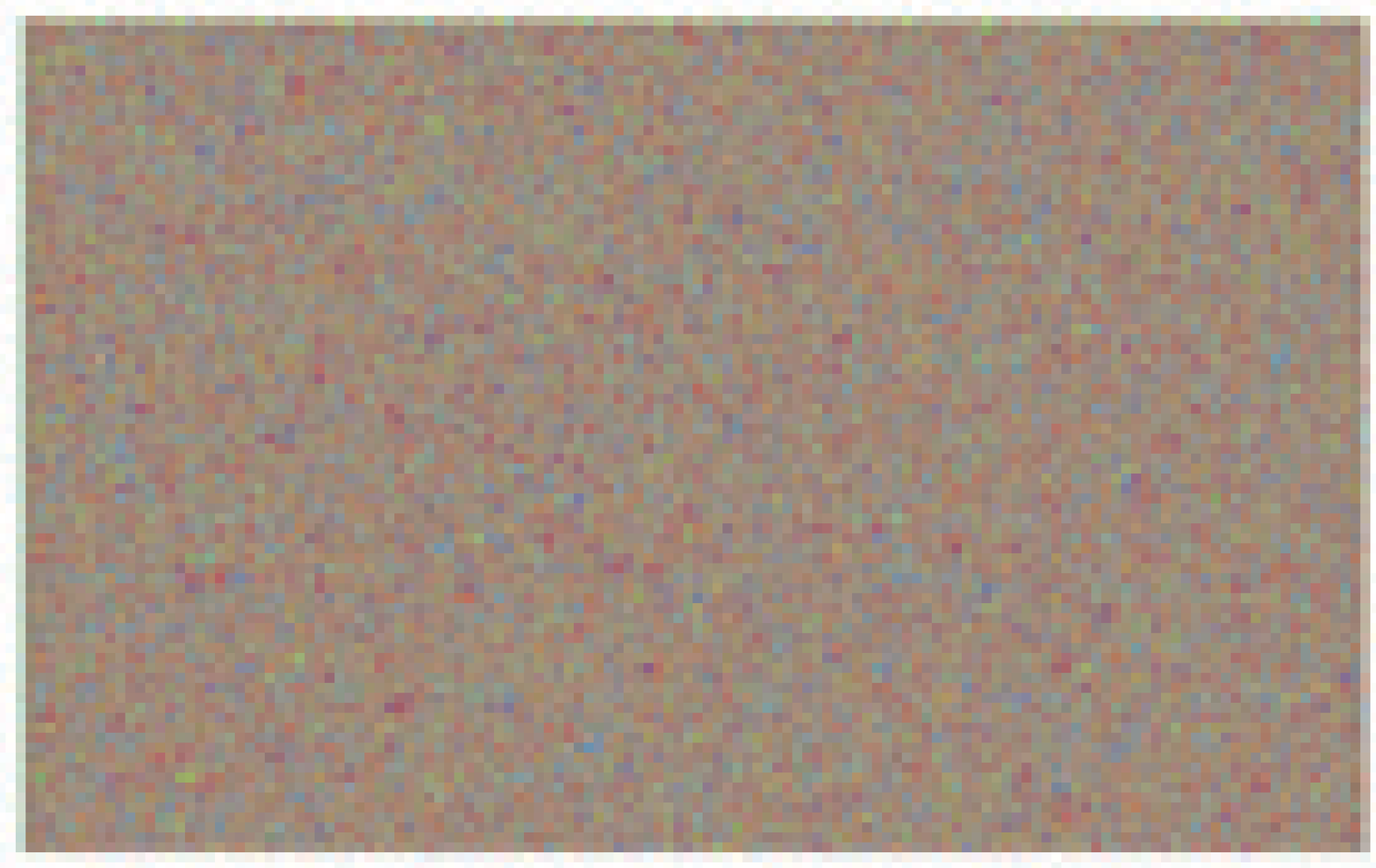} \\
a. & b.\\

 \vspace{-0.5cm}
\end{tabular}
\end{center}
\caption{The images obtained by consecutive $e$--transformations
with the quasigroups of order 4 with lexicographic numbers 46 and
47. The transformations are done on a periodic string
$01230123\ldots 0123$ with the length 600 and with the leader 0.}
 \label{Q4647}
\end{table}

In \cite{MarkovskiFSE2005} one can find the same classification
but instead of terms ``fractal'' and ``non-fractal'' classes of
quasigroups they are named by an other property of them - a class
of linear and a class of exponential quasigroups. In the same
paper it is mentioned that when the order of quasigroup increases,
the number of fractal (linear) quasigroups decreases exponentially
compared to the number of non-fractal quasigroups. An additional
classification that is close to the fractal -- non-fractal
classification can be found in \cite{MarkovskiClassification} and
an excellent survey for many types of classifications of
quasigroups is done in \cite{Dimitrova}.

It is really amazing how our experimental findings about the
fractal -- non-fractal classification of quasigroups comply with
the classification of quasigroups that give one-way permutations
and one-way functions. An open problem is to investigate the
relation between these two classifications. Here even without
precise definition of what ``fractal'' quasigroup would mean, we
just give the following conjecture:

\begin{conjecture}
The classes of fractal quasigroups and quasigroups for which there
is a permutation
$E_{\mathbf{L}\overline{A}\overline{A},P(N)+2N}(A)$ coincide.
\end{conjecture}

\section{Some comparative analysis for the quasigroup one-way functions}
In this section we would like to set the following convention: For
a random oracle in the sense of Rudich and Impagliazzo works on
one-way functions (\cite{Ru88}, \cite{ImRu89}), we will take any
quasigroup $(Q,*)$ of order $s$ together with the family
$\mathcal{Q}_N$ of one-way functions that can be defined by that
quasigroup.

Rudich in his PhD thesis \cite{Ru88}, based on a combinatorial
conjecture (which was proved in 2000 by Kahn, Saks and Smith in
\cite{KSS00}) concluded that there exist oracles for which there
exist one-way functions, but there are no one-way permutations.
That is in perfect compliance with our case of quasigroup one-way
functions. If the oracle (quasigroup) is non-fractal, our
Conjecture \ref{NFConj} says that there are no strings
$\mathbf{L}_{Q,I_N,P(N)}$ such that the function
$E_{\mathbf{L}\overline{A}\overline{A},P(N)+2N}(A)$ is a
permutation.

Impaliazzo and Rudich in \cite{ImRu89} showed that ``There exist
an oracle relative to which a strongly one-way permutation exists,
but secure secret-key agreement is impossible.'' That is again in
compliance with quasigroup one-way functions. Namely, since
quasigroup one-way functions rely on combinatorial characteristics
of the quasigroups, in general there are no evident ``shortcuts''
and properties that will define a trapdoor function, that will
enable secure secret-key agreement.

Quasigroup one-way functions are strong one-way functions i.e.
there is only a small set of values on which they can be inverted
in polynomial time. Thus, security amplification of a weak one-way
function by an iterative process, that was established as a very
useful technique in the work of Yao in 1982 \cite{Yao82} is not
necessary for quasigroup one-way functions. This means that the
speed of computation of quasigroup one-way functions can be very
high. Additionally, since the computations are done consecutively,
they can be parallelized in a pipeline, and then the computation
of the function can be done in time $O(P(N))$. Some initial
applications of quasigroup one-way functions and their properties
to be easily parallelized are already done in definition of the
stream cipher Edon80 \cite{Edon80}. In that stream cipher the
IVSetup procedure is in fact a sort of quasigroup one-way
function.

From Theorem 2 it follows that quasigroup one-way functions can
achieve the level of security of $2^n$ attempts to invert the
function with the length of the input being $n$ bits. That is most
efficient construction as far as we know compared to other
candidate one-way functions that require from $2n$ to $10n$ input
bits to reach the security level of $2^n$.

The last property of quasigroup one-way functions that we want to
mention in this paper, and that is similar to the properties that
have been already found in other one-way functions is the property
of a one-way function to be regular i.e. that have equal number of
inversions on every point of their codomain. Namely, in
\cite{GoldreichSecurityAmplif} and \cite{Goldreich11OneWay}
techniques for obtaining 1--1 one-way functions are proposed if
the one-way function is regular. In our numerous experiments,
every time when we have used fractal quasigroup, the obtained
one-way functions were either permutations or regular ones. The
example that we show on Figure \ref{ExampleGeneral}b. is an
example of a regular function, where every point of its codomain
has exactly two inversions. It would be a challenging task to
apply the same techniques to quasigroup one-way functions.

\section{Conclusions and further directions}

In this paper we have given a formal definition and construction
of a new family of one-way functions and one-way permutations.
They are based on quasigroup string transformations, and have
numerous interesting properties. By some of those properties (such
as speed of computation, security level of inversion) they
outperform all currently known candidate one-way functions.

Many of our results concerning these functions are still
experimental, and thus we have set up several conjectures about
them. We hope that the intriguing experimental results mentioned
in this paper about the new family of one-way functions will be
interesting enough to attract attention of other researchers.


\newpage
%
%

\end{document}